\documentclass[prb,amssymb,twocolumn,showpacs,superscriptaddress,floatfix]{revtex4}

\usepackage{graphicx}
\usepackage{bm}
\usepackage{mathptmx}

\begin{document}

\title{Universality in three-dimensional Ising spin glasses: Nonequilibrium dynamics \\ from Monte Carlo simulations}

\author{F. Rom\'a}
\affiliation{Departamento de F\'{\i}sica, Universidad Nacional de San Luis. INFAP, CONICET. Chacabuco 917, D5700BWS San Luis, Argentina.}

\begin{abstract}
The non-equilibrium dynamics of the three-dimensional Edwards-Anderson spin-glass model with different bond distributions is investigated by means of Monte Carlo simulation.  A numerical method is used to determine the critical temperature and the scaling exponents of the correlation and the integrated response functions.  The results obtained agree with those  calculated in equilibrium simulations and suggest that the universality class does not depend on the exact form of the bond distribution. 
\end{abstract}

	 %Spin glasses and other random magnets

\date{\today}

\maketitle

The critical behavior of spin glasses is a fundamental subject of interest in the statistical mechanics of disordered and frustrated systems.  The real nature of the phase transition is still under discussion and the controversy about different related topics has not been resolved satisfactorily. In this context, it has been suggested that the basic universality rule which state that the critical exponents depend on the dimension of space, the number of order parameter components and the range of interactions, does not hold in spin glasses. \cite{Bernardi1996,Mari1999,Henkel2005,Pleimling2005}  In particular, by means of a technique that combines equilibrium and dynamic measurements, it was shown that the universality class of the three-dimensional (3D) Edwards-Anderson spin-glass model \cite{EA}, depends on the exact form of the interaction distribution function.  On the other hand, for several choices of bond distributions, the finite-size scaling analysis of the Binder cumulant, the correlation length and susceptibility calculated in equilibrium simulations, suggests that this model obeys universality. \cite{Katzgraber2006} In addition, there is not satisfactory agreement between the critical temperatures $T_c$ calculated by non-equilibrium and equilibrium techniques.   

In this work we propose a non-equilibrium method to determine the $T_c$ and the scaling exponents of the correlation and the response functions.  We use this technique to study the 3D Edwards-Anderson spin-glass model with three different bond distributions.  The values the $T_c$ that we obtain agree with those calculated in equilibrium simulations.  Also, we found that the critical exponents are very close to each other, suggesting that the most probable scenario is that the universality class does not depend on the bond distribution form. 

The simplest non-equilibrium method to determine the critical point is based on the temporal relaxation of the order parameter. \cite{Ozeki2007}  First, at time $t=0$ the system is prepared in a fully ordered state and the dynamics is simulated with a standard Monte Carlo algorithm.  $T_c$ is estimated as the temperature at which, in the asymptotic regime, the order parameter follows a power law. This method is appropriate to study a wide variety of systems.  However, although the slow dynamics present in disordered and frustrated systems favors the application of non-equilibrium techniques, in general for spin glasses these do not allow an accurate determination of $T_c$.  The major problem is that in simulations, different quantities seem to decay by a power law in a relatively wide interval of temperatures.  To overcome this difficulty, additional scaling analysis of order parameter or susceptibility have been used to improve the resolution of these methods. \cite{Bernardi1996,Ozeki2001,Nakamura2003} In addition, recently it has been proposed a different procedure based on the divergence of the relaxation time approaching the critical point. \cite{Lippiello2010}  

Now, we will describe the non-equilibrium method proposed in this work. A typical protocol is used, which consists on a quench at $t=0$ from a disordered state ($T \to \infty$) to a low temperature $T$. From this initial condition we simulate the system with a Model A dynamics. \cite{Hohenberg1977}  Then, different two-time quantities are calculated which depend on both the waiting time $t_w$, when the measurement begins, and a given time $t>t_w$.  In particular, for a system formed by $N$ Ising spins, we determine the two-time autocorrelation function defined as 
\begin{equation}
C(t,t_w) = \frac{1}{N} {\left[ \sum_{i=1}^N \langle \sigma_i(t) \sigma_i(t_w) \rangle  \right]}, \label{corr1}
\end{equation}
where $\sigma_i = \pm 1$, $\langle ... \rangle$ indicates an average over different thermal histories (different initial configurations and realizations of the thermal noise), and $[...]$ represents an average over different disordered samples.  The scaling relation for $C$ is 
\begin{equation}
C(t,t_w) = {t_w}^{-b} f_c (t/t_w),  \label{corr2}
\end{equation}
where $b$ is a non-equilibrium exponent. \cite{Pleimling2005,Godreche2002}  

On the other hand, the corresponding two-time linear autoresponse function is $R(t,t_w)=\frac{1}{N} {\left[ \sum_{i=1}^N \delta \langle \sigma_i(t) \rangle/ \delta h_i(t_w) \right]}$, where $h_i(t_w)$ is a time-dependent conjugated external field.  We calculate in simulation the quantity   
\begin{equation}
\rho(t,t_w) = T \int_0^{t_w} du R(t,u), \label{resp1}
\end{equation}
which is the integrated response when switching on the perturbation only for times $t < t_w$.  The scaling relation for $\rho$ is  
\begin{equation}
\rho(t,t_w) = {t_w}^{-a} f_{\rho} (t/t_w),  \label{resp2}
\end{equation}
where $a$ is another non-equilibrium exponent.  For critical systems we have that $b=a$ and, for $t_w \ll t-t_w $, the scaling functions should follow a power-law decay, i. e., $f_c \sim (t/t_w)^{-\lambda_c/z_c}$ and $f_{\rho} \sim (t/t_w)^{-\lambda_c/z_c}$, where $\lambda_c$ is the autocorrelation exponent and $z_c$ is the dynamical critical exponent. \cite{Pleimling2005,Godreche2002} 

The method proposed in this work consists in to calculate $C$ and $\rho$ for different values of $T$ and $t_w$ and, by means of a data collapse analysis, to determine the exponents $b$ and $a$.  Then, $T_c$ is identified as the temperature for which the condition $b=a$ is satisfied. As we shall see, this strategy will allow us to carry out a precise determination of the critical point.          

First, in order to validate the method, we study the two-dimensional (2D) ferromagnetic Ising model on the square lattice, for which $T_c=2/\ln(1+\sqrt{2})$. The Hamiltonian is $H=-\sum_{(i,j)} J \sigma_{i} \sigma_{j}$, where the sum run over the nearest-neighbor pairs and $J=1$.  A large lattice of linear size $L=300$ ($N=L^2$) with fully periodic boundary conditions was simulated using a standard Glauber dynamics and, for each temperature studied, the averages were calculated over $5000$ independent thermal histories.  The correlation was calculated as usually, but the integrated response to an infinitesimal magnetic field was determined using the algorithm proposed in Ref.~\onlinecite{Chatelain2003}.  This is very important to obtain a reliable value of the exponent $a$, making possible the realization of the method studied here.  

\begin{figure}
\includegraphics[width=\linewidth,clip=true]{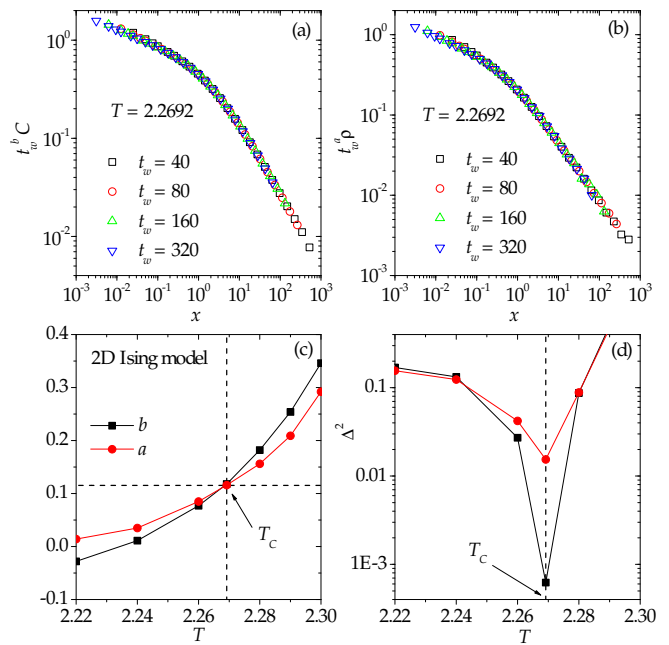}
\caption{\label{figure1} (Color online) The 2D ferromagnetic Ising model. Panels (a) and (b) show, respectively, the data collapse of the correlation and integrated response at $T=2.2692$ for different $t_w$ as indicated.  (c) The values of non-equilibrium exponents $b$ and $a$, and (d) the quality of the collapses for different temperatures.}
\end{figure}

Figures~\ref{figure1} (a) and \ref{figure1} (b) show, respectively, the data collapse of the correlation and the integrated response for  different $t_w$ and temperature $T = 2.2692 \approx T_c$, plotted as function of the variable $x=t/t_w -1$ (we have used $x$ instead of $t/t_w$ but this choice has no consequence on the asymptotic behavior).  The best data collapse was obtained by minimizing the sum of squared differences between all pairs of curves within a given range of $x$.  For the range $x\ge x_0= 1$, the exponents calculated in this way are $b=0.118$ and $a=0.116$.  On other hand, if a common exponent $c=b=a$ it is used to collapse simultaneously both sets of curves, we obtain a value of $c=0.117$.  These numbers are very close to expected value for the Ising model, $b=(d-2+\eta)/z_c \approx 0.115$, where $d=2$ is the dimension of the space, $\eta=1/4$ is the well-known static critical exponent associated to the pair correlation function and $z_c=2.1667$. \cite{Godreche2002,Nightingale2000}  As shown in Fig.~\ref{figure1} (c), repeating this procedure for others  temperatures the values of $b$ and $a$ cross at $T_c$, where an optimal collapse is obtained [see Fig.~\ref{figure1} (d), where $\Delta^2$ is equal to the sum of the squared differences between all pairs of curves].       
 
The previous example shows that the critical point can be located by looking for the temperature at which the condition $b=a$ is fulfilled.  This temperature can be regarded as $T_c$ and the $c$ value as a reasonable estimate of the non-equilibrium critical exponent $b$ (or $a$). However, notice that for temperatures above or below $T_c$, only pseudo-exponents are obtained with this simple method.  Close to (but not at) $T_c$, to calculate (when possible) the true non-equilibrium exponents of a given system, long-time simulations are necessary and maybe, either another scaling relations or a separation of the correlation and the response functions in their corresponding stationary and aging terms are required. \cite{Lippiello2006})  

We now consider the 3D Edwards-Anderson spin-glass model whose Hamiltonian is, $H = - \sum_{(i,j)} J_{ij} \sigma_{i} \sigma_{j}$, where the sum runs over the nearest neighbors of a cubic lattice with fully periodic boundary conditions. The bonds $J_{ij}$'s are independent random variables drawn from a given distribution $P(J_{i,j})$ with mean zero and variance one. As Ref.~\onlinecite{Pleimling2005}, we concentrate on three distributions: the bimodal $P_B(J_{i,j})=\left[\delta(J_{i,j}-1)+\delta(J_{i,j}+1)\right]/2$, the Gaussian $P_G(J_{i,j})=\exp( J_{i,j}^2/2)/\sqrt{2 \pi}$, and the Laplacian $P_L(J_{i,j})=\exp(-\sqrt{2}|J_{i,j}|)/\sqrt{2}$ distribution.  In order to avoid confusions we will denominate, respectively,  EAB, EAG and EAL models to each one of these versions of the Edwards-Anderson model.   Lattices of linear size $L=50$ ($N=L^3$) were simulated using a standard single-spin Glauber dynamics and six values of $t_w=50$, $100$, $200$, $400$, $800$, and $1600$ were used.  The integrated response function was calculated as before for an infinitesimal external field. The disorder average was performed over $3000$ to $5000$ different samples for each temperature.  Because it is expected that the scaling relations, Eqs.~(\ref{corr2}) and (\ref{resp2}) are valid for large values of the waiting time, we have studied different range of $x$ and systematically we have discarded the curves with smallest $t_w$. 

\begin{figure}[t]
\includegraphics[width=8cm,clip=true]{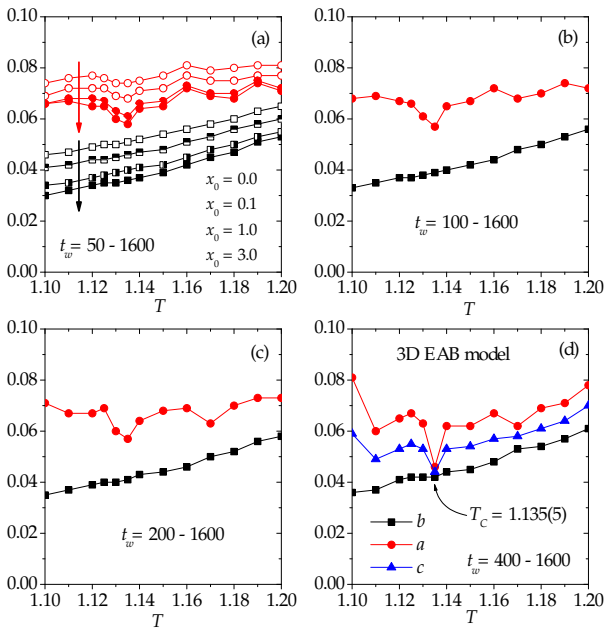}
\caption{\label{figure2} (Color online) Non-equilibrium exponents as function of temperature for the 3D EAB model, obtained for different ranges of $t_w$ as indicated.  Curves in panel (a) were calculated for different values of $x_0$ (arrows indicate how the curves change with increasing $x_0$), while in (b), (c) and (d) $x_0=3$ was chosen.}
\end{figure}

\begin{figure}[t]
\includegraphics[width=8cm,clip=true]{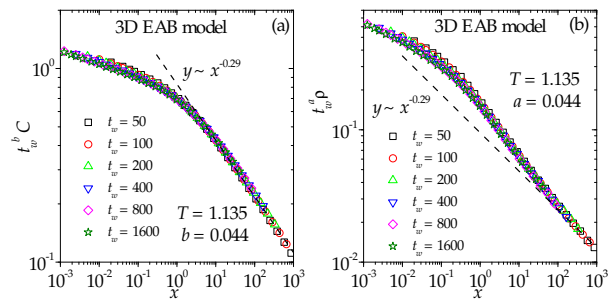}
\caption{\label{figure3} (Color online)  Data collapsing of (a) the correlation and (b) the integrated response functions, for the 3D EAB model at $T=1.135$.  The value $b=a=0.044$ was used.}
\end{figure}

Figure~\ref{figure2} shows the results for the 3D EAB model.  In panel (a) we can see the exponents $b$ and $a$ obtained by the present method for different ranges of $x \ge x_0$, where all curves of $C$ and $\rho$ with $t_w$ between $50$ and $1600$ were considered.  Because the data agree reasonably well each other for $x_0 > 3$, from now on we will use  $x_0=3$. Notice that the condition $b=a$ is not fulfilled at any temperature.  However, after discarding the data for $t_w=50$ and next for $t_w=100$, the Figs.~\ref{figure2} (b) and \ref{figure2} (c) show that the curves begin to come closer together. 
Finally, in panel (d) for the last three $t_w$, we observe that the condition $b=a$ is approximately fulfilled for $T=1.135$, where we obtain $b=0.042$, $a=0.046$ and $c=0.044$ (let us notice that it is not necessary that the curves cross to identify the critical point).  As this behavior is observed among $T=1.13$ and $T=1.14$, we conclude that $T_c=1.135(5)$ for the 3D EAB model.  This value is very close to those obtained in the equilibrium simulations, e. g., $T_c=1.120(4)$ (Ref.~\onlinecite{Katzgraber2006}) and $T_c=1.109(10)$, \cite{Hasenbusch2008} but is slightly lower than $T_c=1.17(4)$ (Ref. \onlinecite{Nakamura2003}) and $T_c=1.19(1)$, \cite{Pleimling2005} two critical temperatures calculated from non-equilibrium simulations.  Nevertheless, notice in Fig.~\ref{figure2} the tendency of the curves to merge at $T=1.17$. Although a great number of samples were calculated, it was not possible to show that the condition $b=a$ can be fulfilled at this temperature. In addition, Fig.~\ref{figure3} shows the data collapse of the correlation and the integrated response functions, where we have used $b=a=0.044$.  

\begin{figure}[b]
\includegraphics[width=8cm,clip=true]{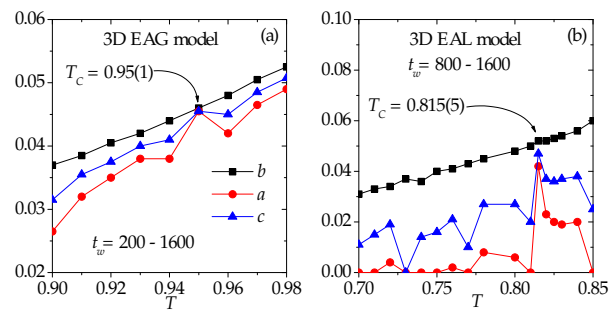}
\caption{\label{figure4} (Color online) Non-equilibrium exponents as function of temperature for (a) the 3D EAG and (b) the 3D EAL models, obtained for different ranges of $t_w$ as indicated. }
\end{figure}

We have used the same protocol to study the others spin-glass models. Figure~\ref{figure4} (a) shows that for the 3D EAG model the condition $b=a$ is approximately satisfied at $T_c=0.95(1)$, where we determine that $b=0.046$, $a=0.0455$ and $c=0.0455$.  This condition is accomplished for $t_w \ge 200$.  Again, this critical temperature agree very well with the value $T_c=0.951(9)$ obtained in equilibrium simulations \cite{Katzgraber2006} but, in this case, is slightly higher than $T_c=0.92(1)$, the value reported in Ref.~\onlinecite{Pleimling2005}.  

On the other hand, Fig.~\ref{figure4} (b) shows that for the EAL model is needed to discard the first four $t_w$ to obtain a reliable value of $T_c=0.815(5)$.  At this temperature we determine that $b=0.052$, $a=0.042$, and $c=0.047$.  The difference between these exponents is larger than the measure in previous models.  Probably, this is due to the fact that the $T_c$ for this model is very low and bigger values of $t$ and $t_w$ need to be reached. Nevertheless, Fig.~\ref{figure4} (b) shows strong evidence that this value corresponds to the true critical temperature which, as before, is found to be slightly higher than $T_c=0.72(2)$, the value reported previously. \cite{Pleimling2005}               

\begin{figure}[t] 
\includegraphics[width=8cm,clip=true]{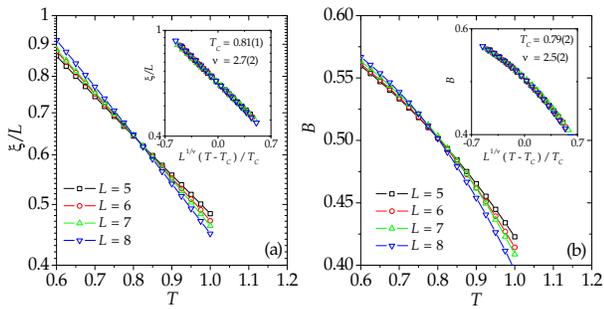}
\caption{\label{figure5} (Color online) (a) Correlation length and (b) Binder cumulant as function of $T$ for the 3D EAL model. Insets: data collapsing.  }
\end{figure}

For comparison, we have also calculated $T_c$ for the 3D EAL model in an equilibrium simulation, using a parallel tempering algorithm. \cite{Geyer1991,Hukushima1996}  To reach equilibrium between $T=1.0$ and $T=0.6$, it was necessary to simulate $17$ replicas of the system and a number of Monte Carlo sweeps of $2^n$ with $n=2L+8$.   At least up to $5 \times 10^4$ samples for each lattice size were necessary to obtain a reliable disorder average.  Due that the number of sweeps grow very fast with $L$, only lattices up to $L=8$ could be studied (probably this is due to the fact that $T_c$ is lower than in the previous models). Figure~\ref{figure5}  shows that the correlation length $\xi/L$ (Ref. \onlinecite{Palassini1999}) and the Binder cumulant $B$ (Ref. \onlinecite{Binder1981}) cross at, respectively, $T_c=0.81(1)$ and $T_c=0.79(2)$.  Both values are compatible with the $T_c$ obtained with our non-equilibrium method.

The quantities estimates in this work for the three spin-glass models are shown in Table~\ref{table}.  We report the values of exponent $c$ (our best estimate of the true values of $b$ and $a$), for which the error bar was determined taking the values of $b$ and $a$ as, respectively, the upper and the lower bounds of $c$.  As we discuss before, the critical temperatures that we have calculated here, agree very well with those obtained in equilibrium simulations.  However, these $T_c$ and the corresponding non-equilibrium critical exponents, differ from values reported in the literature: $T_c=1.19(1)$ and $b=0.056(3)$ for the EAB, $T_c=0.92(1)$ and $b=0.043(1)$ for the EAG, and $T_c=0.72(2)$ and $b=0.032(2)$ for the EAL model. \cite{Henkel2005,Pleimling2005}  It is important to point out that at these same temperatures we obtain similar values of $b$: $b=0.057$, $b=0.0405$, and $b=0.034$ for, respectively, the EAB, the EAG and the EAL models. However, contrary to previous reports, we do not identify these temperatures with the $T_c$ of each system.   

\begin{table}[t]
\caption{\label{table} Quantities calculated in this work.}
\begin{tabular}{cccc}
\hline
\hline
Parameter  			&    EAB      &    EAG     &   EAL   \\
\hline
$T_c$      			&  $1.135(5)$ & $0.95(1)$  &  $0.815(5)$     \\
$c$        			&  $0.044(2)$ & $0.0455(5)$&  $0.047(5)$     \\
$\lambda_c/z_c$   &  $0.29(1)$  & $0.29(1)$  &  $0.254(2)$     \\
$X_{\infty}$      &  $0.09(4)$  & $0.09(1)$  &  $0.04(1)$      \\
\hline
\hline
\end{tabular}
\end{table}

In addition, the values of exponent $c$ that we have obtained are very similar to each other, and they are also quite close to $b \approx 0.0464$, the calculated value from the relation $b=(d-2+\eta)/2z_c$, \cite{Ogielski1985} where $d=3$ and we have used $\eta=-0.375$ (Ref. \onlinecite{Hasenbusch2008}) and $z_c=6.74$. \cite{Katzgraber2005} As it has been shown previously in equilibrium simulations, \cite{Katzgraber2006} these new results suggests that the universality class of the 3D Edwards-Anderson spin-glass model does not depend on the exact form of the bond distribution.  We can even notice that the correlation-length exponents for the 3D EAL model, $\nu=2.7(2)$ and $\nu=2.5(2)$ that we have calculated from, respectively, the data collapse of the correlation length and the Binder cumulant (see insets in Fig.~\ref{figure5}), agree very well with those obtained previously for the 3D EAB and the 3D EAG models. \cite{Katzgraber2006,Hasenbusch2008} 

On the other hand, the assumption of that the universality is violated, it is also based on quantities such as $\lambda_c/z_c$ and $X_\infty = \lim_{t_w\rightarrow\infty} \lim_{t\rightarrow\infty} \rho(t,t_w) / C(t,t_w)$, the critical fluctuation-dissipation ratio. \cite{Cugliandolo1994}  Our simulations show that for the 3D EAB and the 3D EAG models, the calculated values of $\lambda_c/z_c$ and $X_\infty$ agree, but differs from the corresponding one for the 3D EAL model (see Table~\ref{table}).  Although both quantities are believed to take universal values, \cite{Godreche2000} recently this conjecture has been questioned, showing that models belonging to the same universality class at equilibrium, have different values of $\lambda_c/z_c$ or $X_\infty$. \cite{Chatelain2004}  To determine if these results are evidence of the non-universal character of $\lambda_c/z_c$ and $X_\infty$, or of the existence of different non-equilibrium universality classes, further investigations are required. \cite{Calabrese2005} 

In summary, we have proposed a non-equilibrium method to determine the $T_c$ and the scaling exponents of the correlation and the integrated response functions.  We apply this technique to the 3D Edwards-Anderson spin-glass model with three different bond distributions.  The values of $T_c$ that we obtain agree with those calculated in the equilibrium simulations.  As the values of the exponent $c$ are very close to each other, we conclude that for the 3D spin glasses the most probable scenario is that the universality class does not depend on the exact form of bond distribution.   

I thank S. Bustingorry, A. B. Kolton, D. Dom\'{\i}nguez, and  R. Garc\'{\i}a-Garc\'{\i}a for useful discussions.  This work was supported in part by CONICET and the National Agency of Scientific and Technological Promotion
(Argentina) under Projects 33328 PICT-2005 and 2185 PICT-2007.

%.............................................................................................

\end{document}